\documentclass[prb,amsmath,showpacs,preprint,floatfix]{revtex4}
\usepackage{graphicx}
\usepackage{amssymb}

\begin{document}

\title{Correlated-electron description of the photophysics of thin films of $\pi$-conjugated polymers}

\author{D. Psiachos}
\affiliation{Department of Physics, University of Arizona
Tucson, AZ 85721}
\author{S. Mazumdar}
\affiliation{Department of Physics, University of Arizona
Tucson, AZ 85721}
\date{\today}
\begin{abstract}
We extend Mulliken's theory of ground state charge transfer in a donor-acceptor
complex to excited state charge transfer between pairs of
identical $\pi$-conjugated oligomers,
one of which is in the optically excited state and the other in the ground state,
leading to the formation of a charge-transfer exciton. 
Within 
our theory,
optical absorptions from the charge-transfer exciton should include a low energy
intermolecular charge-transfer excitation, as well as distinct intramolecular excitations
from both the neutral delocalized exciton component and the Coulombically bound
polaron-pair component of the charge-transfer exciton. We report high order
configuration-interaction calculations for pairs of oligomers of 
poly-paraphenylenevinylene (PPV) that go beyond our previous single configuration-interaction
calculation and find all five excited state absorptions predicted using heuristic
arguments based on the Mulliken concept. Our calculated excited state absorption
spectrum exhibits strong qualitative agreement with the complete wavelength-dependent
ultrafast photoinduced absorption in films of PPV derivatives, suggesting that a 
significant fraction of the photoinduced absorption here is from the charge-transfer
exciton. We make detailed comparisons to experiments, and a testable experimental
prediction.

\end{abstract}
\maketitle

\section{Introduction}

It is now generally recognized that interchain interactions play a strong role in the photophysics of
thin films of $\pi$-conjugated polymers (PCPs). 
\cite{Rothberg06,Arkhipov04,Schwartz03,Conwell06,Samuel95,Samuel98,Schweitzer99,Hertel01,Ho01,Lim02,Brown03,Koren03,Clark07,Sheng07,Singh08,Zhang07,Morandi09} While the primary photoexcitation in dilute solutions of PCPs,
which contain widely separated single strands,
is to the optical exciton, interchain species other than the intrachain optical exciton are generated in films.
Since the original demonstration of interchain species in films of 
poly(2-methoxy,5-(2$^{\prime}$-ethyl-hexyloxy)1,4 paraphenylenevinylene) (MEH-PPV) by Rothberg and collaborators,
\cite{Rothberg06} a variety of complex optical phenomena in many other PCPs have been ascribed to interchain
interactions.  Experiments that indicate the formation of interchain species upon
photoexcitation include, (i) photoluminescence (PL) at longer wavelength relative to solutions, with smaller
quantum efficiency, \cite{Rothberg06,Arkhipov04,Schwartz03,Conwell06,Samuel95,Samuel98,Clark07} (ii) delayed 
PL lasting until milliseconds, and its quenching by
an electric field, \cite{Rothberg06,Arkhipov04,Schweitzer99,Hertel01} and (iii) {\it additional} photoinduced absorptions (PAs) in ultrafast spectroscopy of 
films over and above those observed in solutions. \cite{Rothberg06,Sheng07,Singh08} 

Three different kinds of interchain species can in principle occur in PCPs. These include,
(i) neutral exciton delocalized over multiple chains, as in an aggregate, \cite{Clark07,Cornil98,Spano05,Barford07}
(ii) completely ionic
polaron-pair, with positively and negatively charged neighboring chains that are Coulombically bound, 
\cite{Rothberg06,Arkhipov04,Conwell06},
and (iii) the charge-transfer (CT) exciton, which is a quantum mechanical superposition of the delocalized neutral exciton and
the ionic polaron-pair that is obtained
in the presence of significant interchain electron hopping. \cite{Conwell06,Wang08} All three 
kinds of interchain species have been proposed to explain experiments in real systems.
PL at wavelengths longer than in solutions
with weaker quantum efficiency can be explained using the H-aggregate scenario, 
wherein absorption is to the highest state in the exciton band and emission occurs from the lowest state in
the band. \cite{Clark07,Spano05}
On the other hand, PA in films was originally thought to be from a polaron-pair state. \cite{Rothberg06,Conwell06}
Very recently, Wang {\it et al.} \cite{Wang08} have claimed that the detailed features of the PA in films
\cite{Sheng07,Singh08} 
can be understood only within the CT exciton picture. 
According to Wang {\it et al.} a significant fraction 
of both PL and PA occur from the CT exciton.
The calculations of the PA in Reference \onlinecite{Wang08} were based on the single configuration-interaction (SCI)
approximation, and were able to reproduce only a limited low energy region of the experimental PA spectra.
\cite{Sheng07,Singh08} In the present work we go beyond SCI and perform quadruple CI and multiple 
reference singles and double CI (MRSDCI) calculations \cite{Buenker74,Tavan87} of excited state absorptions in interacting pairs of 
PPV oligomers to understand the PA spectra over the complete experimental energy range. Our calculated PA
spectra, originating from the CT exciton, exhibit strong
qualitative agreement with the experimental PA spectra in films of PPV 
derivatives, \cite{Sheng07}
in spite of the 
short lengths of the oligomers
for which the many-body calculations could be performed.

It is important for what follows that we clearly 
state the difference
between PA in solutions and films. Both solutions and films show absorptions labeled PA$_1$ and PA$_2$. The lower
energy PA$_1$ exhibits a peak at $\sim$ 1 eV and the higher energy PA$_2$ occurs at $\sim$ 1.3--1.4 eV. 
Two distinct additional PAs, P$_1$ at $\sim$ 0.4 eV and P$_2$ close to, but distinct from PA$_2$, are seen in films.
\cite{Sheng07,Singh08} 
Whether or not P$_1$ and P$_2$ arise from free polarons (generated by exciton dissociation) or from
a bound interchain species has been controversial in the past 
(see discussions in reference \onlinecite{Wang08}). 
Recent experiments have found similar behavior in heterostructures consisting of donor and acceptor PCPs,
\cite{Morteani04,Sreearunothai06} or of PCPs 
and finite molecular acceptors. \cite{Drori08} 
Identifying the nature of the dominant intermolecular species, and determining their
photophysics is clearly important.

In section II we discuss our theoretical model and the methods we use. In view of the highly correlated
natures of the two-chain wavefunctions, and the complexity of the phenomenon of
excited state charge transfer that is only beginning to be understood, we present in section III
heuristic physical descriptions of all photoexcitations that are possible from the CT
exciton. This discussion is an extension of Mulliken's theory of ground state charge transfer
between a donor and an acceptor
\cite{Mulliken52} to excited state charge transfer between identical pairs of molecules,
one of which is in the optically excited state.
The qualitative arguments make the computational results, presented in section IV, understandable. 
We present our conclusions, along with comparisons to experiments, in section V. 

\section{Theoretical model and methods}

As in Reference \onlinecite{Wang08}, we consider pairs of capped PPV oligomers 4 nm apart, stacked cofacially. 
The sophisticated many-body approaches used here require enormous Hamiltonian bases, \cite{Shukla03} which necessarily
limits us to oligomers that are considerably shorter than those investigated in our
previous SCI calculations. \cite{Wang08} Our calculations here are for pairs of oligomers of length
three and four units each. We refer to the individual oligomers as PPV3 and PPV4, respectively.
Note that with two such
chains these calculations are much larger than the previous QCI and MRSDCI calculations of PA$_1$
and PA$_2$ in single chains. \cite{Shukla03}
Although
the actual arrangements of the chromophores deviate from ideal cofacial stacking in the real systems,
it is believed that this assumption
captures the essential physics of polymer thin films. \cite{Conwell06,Cornil98,Wang08} 
Unlike in \onlinecite{Wang08}, our 
calculations here are only for pairs of oligomers with equal lengths, so that the entire two-chain
structure possesses a center of inversion. This is for computational simplification only and does
not affect our conclusions. As indicated in our previous work, lack of inversion 
symmetry makes the CT
exciton below the optical exciton weakly optically allowed from the ground state, but the
energy locations and strengths of the calculated PAs originating {\it from} this interchain species  
are the same whether or not inversion symmetry exists. \cite{Wang08}

We describe our system within an extended two-chain Pariser-Parr-Pople $\pi$-electron Hamiltonian 
\cite{Pariser53,Pople53}
$H = H_{intra}+H_{inter}$, where $H_{intra}$ and $H_{inter}$ correspond to intra-
and interchain components, respectively. The two components of $H$ are written as,
\begin{equation}
\label{H_intra}
\begin{split}
H_{intra} = -&\sum_{\mu\langle ij \rangle, \sigma}t_{ij}
(c_{\mu,i,\sigma}^\dagger c_{\mu,j,\sigma}+ H.C.) +
U \sum_{\mu,i} n_{\mu,i,\uparrow} n_{\mu,i,\downarrow}\\
&+ \sum_{\mu,i<j} V_{ij} (n_{\mu,i}-1)(n_{\mu,j}-1)
\end{split}
\end{equation}
and,
\begin{equation}
\label{H_inter}
\begin{split}
H_{inter}^{1e} = -t_{\perp}\sum_{\mu <
\mu^{\prime},i,\sigma}(c^{\dagger}_{\mu,i, \sigma}
c_{\mu^{\prime},i,\sigma} + H.C.)  \\
+ \frac{1}{2}\sum_{\mu < \mu^{\prime},i,j} V_{ij}^{\perp}(n_{\mu,i} -1)(n_{\mu^{\prime},j} - 1)
\end{split}
\end{equation}

In the above $c^{\dagger}_{\mu,i,\sigma}$ creates a $\pi$-electron of spin $\sigma$ on
carbon atom $i$ of oligomer $\mu (=1,2)$, $n_{\mu,i,\sigma} =
c^{\dagger}_{\mu,i,\sigma}c_{\mu,i,\sigma}$ is
the number of electrons on atom $i$ of oligomer $\mu$ with spin $\sigma$ and
$n_{\mu,i} = \sum_{\sigma}n_{\mu,i,\sigma}$ is the total number of electrons on atom
$i$ of oligomer $\mu$. We consider standard intrachain nearest neighbor one-electron hopping integrals
$t_{ij}$ = 2.4 eV for phenyl C-C bonds, and 2.2 (2.6) eV for single (double) C-C bonds,
respectively. \cite{Chandross97} 
$U$ and $V_{ij}$ are the
on-site and intrachain intersite Coulomb interactions.
$V_{ij}$ are obtained from a modification of the Ohno parametrization \cite{Ohno64}
\begin{equation}
\label{Vij}
V_{ij}=\frac{U}{\kappa\sqrt{1+0.6117 R_{ij}^2}}
\end{equation}
where $R_{ij}$ is the distance between carbon atoms $i$ and $j$ in
\AA, and $\kappa$ is the background dielectric
constant along the chain direction that takes into account screening effects from
the inner-shell electrons. Within the standard Ohno parametrization, $U=11.26$ eV and $\kappa=1$.
We use $U=8$ eV and $\kappa=2$. The detailed justification of our
parameters has been given elsewhere. \cite{Chandross97} 

As in our previous work, \cite{Wang08} interchain hopping is restricted between the closest pairs
of carbon atoms; we choose $t_{\perp}=0.1$ eV. For $V^\perp_{ij}$, we choose 
the same functional forms as in Eq.~\ref{Vij}, with the same dielectric screening for the PPV oligomer. 
As shown before, \cite{Wang08}
the results of the PA calculations are relatively insensitive to variations of the 
transverse dielectric constant, as long as reasonably realistic values are chosen.

In addition to PA calculations for interacting pairs of neutral chains, we have also calculated
the ground state absorption energies of singly-charged three- and four-unit oligomers, for comparison
to the PA spectra of neutral systems. 
These absorption energies can be strongly affected by electron-phonon interactions, which,
however,
have not been included in our $H_{intra}$. This is primarily because calculating PA within a model that
incorporates both electron-electron and electron-phonon interactions is currently difficult, and 
the comparison would be meaningless if only one of them was calculated with electron-phonon interactions.
We shall nevertheless refer to the charged oligomer absorptions as polaron absorptions.

\section {Charge-transfer excitons and excited state absorption}

We present here a brief review of our earlier work on the CT exciton in interacting 
PCP chains for the sake of completeness. Following this, we present a diagrammatic description of the excited 
state absorptions that are expected for the two-chain system, within an extension of the Mulliken theory
of charge transfer. \cite{Mulliken52} 

Four distinct basis functions, two neutral and two ionic, are relevant for understanding the correlated eigenstates 
near the optical edge of PCPs. \cite{Wang08} 
These are the excited neutral and charged eigenstates of $H_{intra}$ for two independent chains. 
We write the neutral two-chain basis functions
as product functions of the single-chain 1A$_g$ ground state and the 1B$_u$ optical exciton,
viz., $|exc1\rangle = |1B_u \rangle_1 |1A_g\rangle_2$ and 
$|exc2\rangle = |1B_u \rangle_2 |1A_g\rangle_1$, respectively, where the suffixes are chain indices. 
The charged polaron configurations, in their lowest states, 
are similarly written as $|P^+\rangle_1 |P^-\rangle_2$ and $|P^+\rangle_2 |P^-\rangle_1$. For symmetrically placed
chain pairs, and within the SCI, the total Hamiltonian $H$ has four eigenstates: \cite{Wang08} the even parity optical
exciton, $|EX\rangle = {1 \over \sqrt{2}}[|exc1\rangle + |exc2\rangle]$; the even parity polaron-pair state, 
$|PP\rangle_+ = {1 \over \sqrt{2}}[|P^+\rangle_1 |P^-\rangle_2 + |P^+\rangle_2 |P^-\rangle_1]$; and two distinct charge-transfer
excitons, hereafter $|CTX1\rangle$ and $|CTX2\rangle$, that are superpositions of the odd parity 
states ${1 \over \sqrt{2}}[|exc1\rangle$ - $|exc2\rangle]$  and 
${1 \over \sqrt{2}}[|P^+\rangle_1 |P^-\rangle_2 - |P^+\rangle_2 |P^-\rangle_1]$. For the interchain
dielectric screening constant less than or equal to the intrachain dielectric screening, the energy ordering
of the states \cite{Wang08} is $|CTX1\rangle < |EX\rangle < |PP \rangle_+ < |CTX2\rangle$.

Within the Mulliken theory, \cite{Mulliken52} a weakly coupled donor-acceptor system is expected to exhibit
a low energy charge-transfer absorption as well as weakly perturbed molecular absorptions. 
Current pump-probe experiments have very high sensitivity, with $|\Delta T/T| \sim 10^{-4}$ (where $T$ is the
transmission of the probe beam without the pump beam, and $\Delta T$ the difference in the transmission with and
without the pump) \cite{Sheng07}. Under this condition, 
{\it excited state ``molecular absorptions'' from $|CTX1\rangle$ 
will be visible from both the neutral exciton component as well as the ionic polaron-pair component.}
We show this schematically in Fig.~1,
where we have included all possible excited state absorptions from $|CTX1\rangle$.  Here $|Ex_-\rangle$ and 
$|PP_- \rangle$ are
the superpositions of the odd parity exciton and polaron-pair configurations mentioned above; we have
shown only one member of each symmetry-adapted pair in both cases. Each two-chain configuration is shown as a
product state of single-particle configurations, with electrons occupying bonding and antibonding molecular
orbitals (MOs), and spin-singlet bonds between the MOs that are singly occupied by electrons.

\begin{figure}
 \centering
 \includegraphics[clip,width=4.0in]{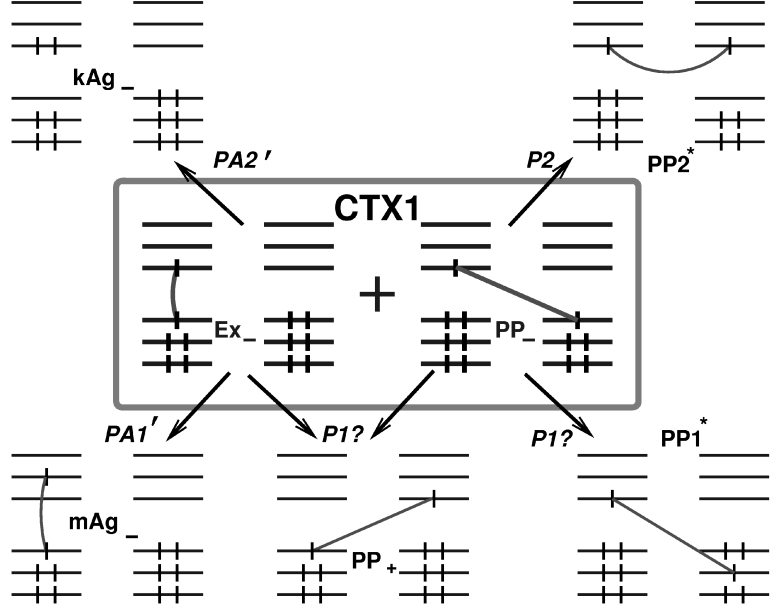}
\caption{Schematic of the PAs from the charge-transfer exciton $|CTX1\rangle$, shown in the box as a
superposition of the neutral exciton state $|Ex_-\rangle$ and the polaron-pair state $|PP_-\rangle$
(see text). 
$|PP_+\rangle$ is reached by low energy charge-transfer excitation; $|mAg_-\rangle$ and
$|kAg_-\rangle$ are reached by 1e-1h and 2e-2h intramolecular excitations from the $|Ex_-\rangle$ component
of $|CTX1\rangle$, giving induced absorptions PA$_1^{\prime}$ and PA$_2^{\prime}$, respectively.
$|PP1^*\rangle$ and $|PP2^*\rangle$ are reached by 1e-1h intramolecular excitations
from the $|PP_-\rangle$ component of $|CTX1\rangle$ (see text). 
}
\end{figure}
The charge-transfer excitation from $|CTX1\rangle$ in Fig.~1 is to the even parity polaron-pair state $|PP\rangle_+$.
This state is reached by intermolecular transfer of an electron from the 
antibonding (bonding) MO of the molecule in the excited (ground) state of $|Ex_-\rangle$ to the antibonding (bonding)
MO of the molecule in the ground state. The polarization of this transition, in contrast
to the intramolecular transitions discussed below, is perpendicular to the plane of the
molecules.

Two different {\it intra}molecular excitations from the $|Ex_-\rangle$ component of $|CTX1\rangle$ 
are indicated in Fig.~1.
The first of these is to a higher energy {\it effective} one electron-one hole (1e-1h) excited state
that we have labeled as the $|mA_{g-}\rangle$. The second is an effective two electron-two hole (2e-2h) 
excitation labeled  $|kA_{g-}\rangle$. 
Corresponding to each of these intramolecular excitations of the two-chain CT exciton
there exist single-chain excited states
that have been extensively discussed in the context of nonlinear absorptions in single chains. \cite{Shukla03,Chandross97}
The single-chain $|mA_g\rangle$ and the  $|kA_g\rangle$ are the final
states corresponding to induced absorptions PA$_1$ and PA$_2$, respectively, from the single-chain $|1B_u\rangle$ exciton. 
\cite{Frolov00,Frolov02}
The minus subscripts in $|mA_{g-}\rangle$ and the $|kA_{g-}\rangle$ imply that as with the $|Ex_-\rangle$, 
the actual wavefunctions 
contain odd parity superpositions
of two-chain configurations. 
The $|mA_{g-}\rangle$ and the  $|kA_{g-}\rangle$ are higher energy CT excitons, and also contain 
excited polaron-pair components that are not explicitly shown in the Figure. 
Indeed, each final state wavefunction in Fig.~1 is a superposition of several dominant configurations, only
one of which is shown explicitly in the Figure. The others are obtained by applications of mirror-plane and
charge-conjugation symmetry operations on the configurations shown.
In the case of the $|mA_{g-}\rangle$ the
exciton and excited polaron-pair contributions to the wavefunction were both demonstrated within the
SCI in our previous work. \cite{Wang08}  

Fig.~1 also indicates intramolecular absorptions from the $|PP_-\rangle$ component of $|CTX1\rangle$. We have labeled 
the dominant final configurations here as $|PP1^*\rangle$ and $|PP2^*\rangle$, which are bound high energy 
excited polaron-pair
configurations. 
For realistic $t_{\perp}$ and
$V_{ij}^{\perp}$, the 
energies of the excited state transitions from the CT exciton to the polaron-pair states 
$|PP1^*\rangle$ and $|PP2^*\rangle$ should be comparable to the ground state absorption energies 
in charged chains, since the former are exclusively from the charged component of $|CTX1\rangle$.

\section{Computational results}

We find from direct computations of excited state spectra that all of the absorptions
indicated in Fig.~1 should indeed be observed.
We report results of QCI and MRSDCI calculations on pairs of PPV3 and PPV4 oligomers.
In our many-body calculations we have retained all innermost one-electron delocalized
bonding and antibonding band levels, and the localized bonding and antibonding bands that occur below and above
these. \cite{Chandross97} The dimension of the QCI Hamiltonian matrix for pairs of PPV3 oligomers was
1,83,3276. In order to obtain convergence in energies to 0.02 eV, the MRSDCI calculations for 
PPV4 were performed with
54 reference configurations. The overall dimension of the Hamiltonian matrix for the two-chain system in this
case was 1,46,8048. These sizes are 
at the limit of our computational capability.

In Figs.~2(a) and (b) we have shown our calculated PAs from $|CTX1\rangle$ for PPV3 and PPV4 oligomers, respectively.
The PA energies are given in units of the energy of the two-chain optical exciton $|EX\rangle$ in each case. 
We have also calculated
the optical absorption energies from the ground states of the charged single chains (PPV3$^+$ and PPV4$^+$). 
The arrows in the figures
denote the charged chain absorption energies, relative to the energy of $|EX\rangle$. The large differences
between the charged-chain absorptions and PAs labeled PP1 and PP2 are primarily due to the finite size effects
associated with the very short system sizes being probed here, although additional contribution to this difference
arises also from interchain Coulomb interactions that influence the energies of the two-chain PAs. 

We have
calculated the PAs from the two-chain optical exciton $|EX\rangle$. 
PAs nearly identical in energy to PA$_1^{\prime}$
and PA$_2^{\prime}$ are found, but the PAs labeled CT, PP1 and PP2 are missing in this case, confirming
the analysis of Fig.~1.  
We have also done detailed wavefunction
analyses of the final states of all the PAs shown in Fig.~2 to confirm that the dominant configurations in each
case are the same as that indicated in the schematic Fig.~1. Thus the final state of the
absorption labeled CT in Fig.~2 is predominantly $|PP_+\rangle$, while the final states
of the absorptions labeled PP1 and PP2 are predominantly superposiitons of the
polaron-pair configurations shown in Fig.~1. 
The results of Fig.~2 go considerably
beyond the SCI studies of reference \onlinecite{Wang08}, which had found only the PA labeled PA$_1^{\prime}$ and
the low energy charge-transfer absorption. While it is expected that inclusion of higher order CI is essential
for detecting the PA labeled PA$_2^{\prime}$ \cite{Shukla03}, it is interesting that this is a requirement also
for the PAs labeled PP1 and PP2, in spite of their predominantly 1e-1h characters.  
A few additional points need to be emphasized. Our earlier SCI calculations were for pairs of 
oligomers 2-3 times longer,\cite{Wang08} and from careful finite-size analysis we concluded that
the features labeled CT and PA$_1^{\prime}$ will continue to be seen for a broad range of
realistic oligomer lengths. While it is not possible to do a similar finite size analysis for
the QCI and MRSDCI calculations we report here, based
on the previous analysis we are convinced that all PAs in Fig.~2 will similarly be seen
at realistic chainlengths. The relative energy separations may, however, be strongly chain-length
dependent. Thus from the nature of the final states $|PP_+\rangle$ and $|PP_1^*\rangle$
in Fig.~1 it is conceivable that the absorptions labeled CT and PP1 in Fig.~2 are energetically
much closer in systems with realistic lengths, especially if electron-phonon interactions play
a weak role. Similarly, the relative locations of PP2 and PA$_2^{\prime}$ may also depend on
system size. Both of these may have experimental consequences, as discussed in the next section.
\begin{figure}
 \centering
 \includegraphics[clip,width=3.5in]{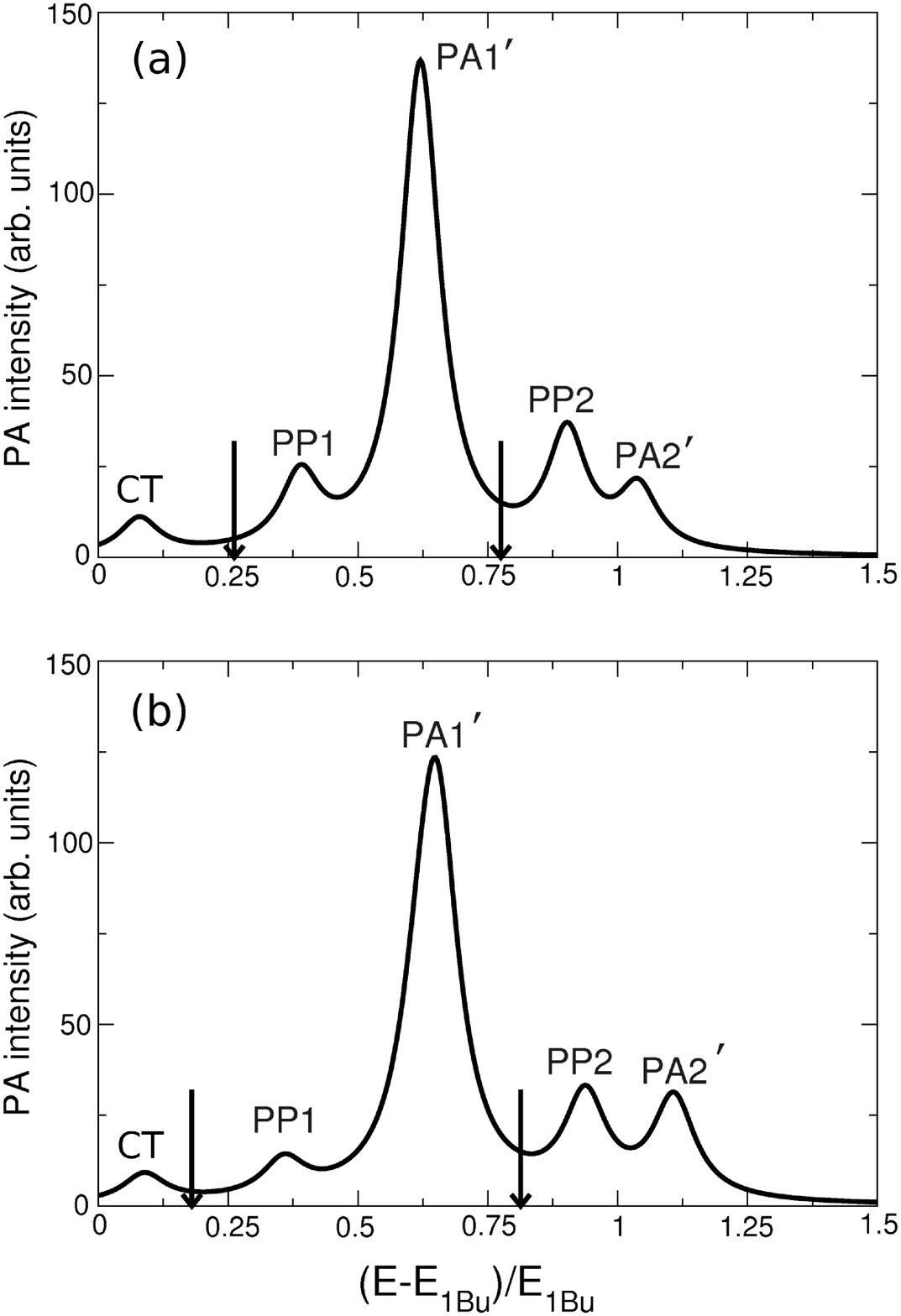}
 \caption{Calculated PAs from $|CTX1\rangle$ in pairs of (a) PPV3, and (b) PPV4 oligomers. 
All energies are relative to that of the neutral optical exciton. Arrows denote energies of
absorption from the ground states of charged chains.}
\end{figure}

\section{Discussions and comparison to experiments}

Within our theory, photoexcitation in films generates predominantly the optical exciton, along with 
small amounts of the CT exciton, which is weakly optically allowed due to disorder. \cite{Wang08}
PL is predominantly from the CT exciton, whose neutral component is the same as the lowest forbidden
exciton state in a H-aggregate. The mechanism of light emission is therefore the same as in theories
emphasizing aggregate formation. \cite{Clark07,Spano05,Barford07} We believe that
PA in films at the earliest times is predominantly from the optical exciton with weak contributions from
the CT exciton, while at later times the PA is predominantly from the CT exciton. 
Recent careful measurements have indicated that the PA labeled PA$_1$ in films exhibit biexponential decay,
which would agree with the two-state scenario. \cite{Vardeny08}
Recent experiments
on pentacene films support this viewpoint. \cite{Marciniak07} Additional support comes from the
absence of free polarons, as noted
in microwave conductivity measurements \cite{Dicker04} and terahertz spectroscopy, \cite{Hendry05}
in spite of the experimentally observed PA bands P$_1$ and P$_2$ that resemble absorption from charged
chains. We believe that an alternate possibility, viz., PL is from the exciton but PA is from a bound polaron-pair,
is precluded by the absence of room temperature infrared-active vibration (IRAV) 
in the recent PA experiment. \cite{Sheng07} Unlike the CT exciton,
the polaron-pair wavefunction is not affected by
interchain charge-transfer and is asymmetric: {\it i.e.}, a specific chain is positively charged
while the other is negatively charged. The counterpart related by symmetry, with the charges on the two chains
reversed, is missing in the polaron-pair, and hence PA even from the bound polaron-pair should  
be accompanied by strong IRAV. Although strong IRAV has been reported by one research 
group, \cite{Miranda01} this results has not been reproduced by any other group.

A key weakness of our theory in Reference \onlinecite{Wang08} was that the PA spectrum could be
calculated only for a limited energy range. This is remedied in the present work.
Our calculated PA spectra of Fig.~2 are to be compared
with the experimental PA spectra of Sheng {\it et al.} (see Figures 1 and 4 in Reference \onlinecite{Sheng07}).
Remarkable qualitative similarities between the calculated and experimental spectra are found, although from
the theoretical spectrum it is difficult to distinguish between the two possible origins of the experimental
P$_1$ absorption.
From our calculations, the P$_1$ absorption observed experimentally can be due to either the low energy 
charge-transfer absorption, or the    
absorption to the $|PP1^*\rangle$ state. We suggest polarization-dependent experimental measurements 
to distinguish between these two
possibilities: the charge-transfer absorption is polarized perpendicular to the molecular planes, while the
absorption labeled PP1 is polarized along the chain direction in the molecular plane. Yet another possibility is
that {\it the experimental P$_1$ absorption is composed of two underlying absorptions.} 
As pointed
out in the previous section, in systems with realistic chain lengths these absorptions will
be proximate in energy and can even be very close. In a recent
experiment performed by Vardeny {\it et al.} on films of PPV derivatives, it has been found that 
the P$_1$ absorption splits into two distinct peaks under hydrostatic pressure. \cite{Vardeny08}
Whether or not these two 
PAs have different origins, as might be expected from our calculations, or they arise from the splitting of 
the same absorption under the experimental conditions, is currently not clear, and further 
experimental work is necessary to clarify this issue. 
 
In conclusion, we have extended the Mulliken theory of charge 
transfer to the case of excited state charge transfer,
to develop a theory of the CT exciton in thin films of $\pi$-conjugated polymers. The central feature of our
theory is that PA in films can occur from both the neutral exciton component of the CT exciton as well as
from the charged bound polaron-pair component. PAs from the exciton components occur at nearly the same
wavelengths as those of the single chains that occur in dilute solutions. PAs from the polaron-pair component occur
at nearly the wavelengths where ground state absorptions of charged chains occur. 
Interestingly, the absorptions from the polaron-pair component of the CT exciton are vanishingly weak in SCI
calculations and become visible only upon including higher order CI.
In addition to these features, the PA spectrum
can exhibit a low energy charge-transfer absorption. Whether or not the low energy P$_1$ absorption seen in
PA measurements \cite{Sheng07} is a superposition of two fundamentally different absorptions is currently
not understood, and will require further experimental work.
There is no such ambiguity
in the case of the absorption we have labeled PP2: this is certainly the origin of the P$_2$ absorption
in the experiment. \cite{Sheng07} Note, however, that the relative energies of PAs labeled PP2 and
PA$_2^{\prime}$ in real systems can be slightly different from those predicted in Fig.~2, since electron-phonon
interactions, ignored in our calculations, can have a strong effect on the energy of the PP2 absorption.
The relative energies of these experimental PAs can perhaps also be dependent on chain length, and hence may be
more strongly conformation-dependent than the lower energy PAs. This may explain the difference in the
conclusions regarding the energy location of P$_2$ between references \onlinecite{Sheng07} and 
\onlinecite{Singh08}. Work is currently in progress to include electron-phonon interactions in our PA calculations
to probe this aspect further.

\begin{acknowledgments}
S.M. acknowledges many stimulating discussions with Professors L. J. Rothberg and Z. V. Vardeny
that led to new insights. 
This work was supported by NSF-DMR-0705163.
\end{acknowledgments}

\end{document}